\documentclass[11pt,a4paper]{article}
\usepackage{jheppub}
\usepackage{graphicx}
\usepackage{epsfig}
\usepackage{amsmath}
\usepackage{amssymb}
\usepackage{amsthm}
\usepackage{ulem}
\usepackage[utf8]{inputenc}
\def\unit{{1\kern-.65ex {\rm l}}}
\def\1{{1\kern-.65ex {\rm l}}}

\def\st{{\widetilde{s}}}
\def\at{{\widetilde{a}}}

\def\ft{{\widetilde{f}}}
\def\gt{{\widetilde{g}}}

\def\wt{{\widetilde{w}}}

\def\CL{{\cal L}}

\title{Searching for BPS Vortices with Nonzero Stress Tensor in Generalized Born-Infeld-Higgs Model}
\author[a,1]{A. Nata Atmaja,\note{Corresponding author.}}

\affiliation[a]{Research Center for Physics \\
Indonesian Institute of Sciences (LIPI)\\
Kompleks PUSPIPTEK Serpong\\
Tangerang 15310, Indonesia.\\}

\emailAdd{ardi002@lipi.go.id}

\abstract{In this article we show that the new BPS equations for vortices, with nonzero diagonal components of the stress tensor, obtained in~\cite{Atmaja:2015lia} for the generalized Maxwell-Higgs model can also be derived using the BPS Lagrangian method developed in~\cite{Atmaja:2015umo}. We add into the original BPS Lagrangian $L_{BPS}=\int dQ$, which is a total derivative term, two additional terms that are proportional to square of the first-derivative of scalar effective field, $f'(r)^2$, and to a function that depends only on the scalar effective field. These additional terms produce additional constraint equations coming from Euler-Lagrange equations of the BPS Lagrangian. We apply this procedure for the generalized Born-Infeld-Higgs model and show that the total static energy, for the corresponding BPS equations, is finite if the scalar potential $V< 2b^2$, with $b$ is the Born-Infeld parameter. We also compute the energy-momentum tensor and show that its diagonal spatial components in radial and angular directions are nonzero. Furthermore we show that the conservation of energy-momentum does not produce new constraint equation. We do the numerical analysis and found that for a large class of solutions the scalar and gauge effetive fields, $f(r)$ and $a(r)$, behave nicely near the origin, but unfortunately they are infinite near the boundary. We suggest that incorporate gravity into the action might resolve this problem and other resolution is by considering BPS vortex in higher dimensional models. We also suggest that the BPS Lagrangian method could be used to find BPS equations for other solitons with nonzero stress tensor.}

\begin{document}
\maketitle
\flushbottom

\section{Introduction}
Vortex is a soliton in two-dimensions, or in general it is solitonic object with the number of co-dimension is two~\cite{Manton:2004tk,Weinberg:2012pjx}. In field theory with three-dimensional spacetime it is a point-like object while in four-dimensional it becomes a string-like object, i.e. vortex strings. In order for vortex to have finite energy, the field theory must be equipped with additional gauge field due to the Derrick's theorem~\cite{Derrick:1964ww}, and so vortex is featured with electromagnetic charges. Vortex finds its applications in many branches of physics. As an example, magnetic vortex of the standard Maxwell-Higgs model (sMH), obtained by Nielsen and Olesen~\cite{Nielsen:1973cs}, correspond to the Type-II superconductor identified by Abrikosov~\cite{Abrikosov:1956sx}. Some other applications of vortex are in Bose-Einstein condensates~\cite{PhysRevLett.83.2498}, in quantum Hall effect~\cite{PhysRevLett.62.82}, including cosmic strings in the early formation of the Universe~\cite{Hindmarsh:1994re}, and many more.

The dynamics of vortex is given by the Euler-Lagrange equations which is a second-order differential equations. These equations in general are non-linear and finding their solutions could be very difficult. In some cases, we may take some limit to the parameters of field theory that could make the Euler-Lagrange equations simpler and easy to solve as shown by Prasad and Sommerfield in the case of monopole and dyon~\cite{Prasad:1975kr}. It turns out that these solutions are also solutions to the first-order differential equations that also solve the Euler-Lagrange equations as shown by Bogomolnyi~\cite{Bogomolny:1975de}. One could obtain these first-order differential equations directly from the static energy density using Bogomolnyi's trick by squaring the energy density, in which the first-order differetial equations are sometimes called Bogomolnyi-Prasad-Sommerfield (BPS) equations. This trick has been used for many solitonic objects including the vortices found in the standard Maxwell-Higgs model~\cite{Nielsen:1973cs}.

The existence of these BPS equations does not only simplify the problem of solving the second-order equations but they also have intimate relation with the supersymmetry extension of the theory~\cite{Weinberg:2012pjx}. Furthermore, in most of the cases, the static energy is bounded from below which is determined fully by topological charge of the theory. The BPS solitons, which are solutions to the BPS equations, saturate this bound of energy thus easy to prove their stability property. However, extracting these BPS equations from the static energy density for general cases is somewhat difficult and tricky. An additional dummy term in the Lagrangian might be needed to execute the Bogomolnyi's trick, with the cost of producing a constraint equation to eliminate this term at the end~\cite{Casana:2015bea,Casana:2015cla}. An attempt was made in~\cite{Bazeia:2007df} by imposing a pressureless condition on the energy-momentum tensor of two-dimensional scalar field theory, with general kinetic term. The kink solutions then can be extracted from this condition. However, in general cases, extracting the BPS equations is also a bit tricky for the higher-dimensional theory. A question was aroused if there is a more rigorous way to obtain these BPS equations. A proposal was given not along ago and it is know as the \textit{On-Shell} method~\cite{Atmaja:2014fha}. The main idea of this method is by introducing auxiliary fields into the (second-order) Euler-Lagrange equations. These auxiliary fields, which would generate the (first-order) BPS equations that are solutions to the Euler-Lagrange equations. For details procedure see~\cite{Atmaja:2014fha, Atmaja:2015lia}. The advantage of this method is that since one works in the Euler-Lagrange equations, whatever Bogomolnyi equations obtained are always solutions to the Euler-Lagrange equations. However, the procedure is a bit tedious and involves solving additional constraint equations though do not depend on coordinate explicitly. Later on, a BPS Lagrangian method was developed in~\cite{Atmaja:2015umo} by introducing a BPS Lagrangian, taking this BPS Lagrangian to be equal to effective Lagrangian of the model, and  then solving it as quadratic equation of first-derivative of all effective fields. We will describe the procedure in more details below. Another a method called First-Order Euler-Lagrange(FOEL) formalism was developed in~\cite{Adam:2016ipc} which is generalization of Bogomolnyi decomposition using a concept of strong necessary condition~\cite{Sokalski:2001wk}.

New BPS equations for vortices, with $C_0\neq 0$, was found in three dimensional generalized Maxwell-Higgs model~\cite{Atmaja:2014fha}. It was shown that these are BPS equations for vortices with nonzero spatial diagonal components of the stress tensor. In the models with four dimensional spacetime, these nonzero diagonal components of the stress tensor have physical meaning as pressure in the corresposding directions. Finding this type of BPS solitons and studying their thermodynamical propesties may have interesting applications in compact stars, or even dark matter~\cite{Adam:2015lpa,Grobov:2015lda}. One might ask if we could find other BPS equations for solitons of this type in other models. In this article, we develop a procedure based on the BPS Lagrangian method of how to obtain this kind of BPS solitons. In particular, we will consider the Born-Infeld extension of the generalized Maxwell-Higgs model, which is generalization of the Born-Infeld-Higgs model~\cite{Shiraishi:1990zi}, that is given by Casana et al. in~\cite{Casana:2015bea}. Initially the new BPS equations for vortices in~\cite{Atmaja:2014fha} were obtained using the \textit{On-Shell} method. However, appying the \textit{On-Shell} method to this model will be very complicated since the Lagrangian has terms inside a square-root. Therefore in this article we will use the BPS Lagrangian method which is shown in~\cite{Atmaja:2015umo} to have a relatively simpler procedure compared to the \textit{On-Shell} method in deriving BPS equations for vortices in the Born-Infeld type of actions. In the section~\ref{sec 2}, we explain in detail about the BPS Lagrangian method and how to use it to find BPS equations for vortices in the standard Maxwell-Higgs model. Next section~\ref{sec 3}, we use the BPS Lagrangian method in the case of generalized Maxwell-Higgs model with the form of BPS Lagrangian, for BPS vortices with nonzero stress tensor, that is motivated by the form of static energy in~\cite{Atmaja:2015lia}. In section~\ref{sec 4}, we then use this form of BPS Lagrangian in the case of generalized Born-Infeld-Higgs model and obtain BPS equations for vortices. In section~\ref{sec 5}, we compute the energy-momentum tensor and show that its radial-radial and angular-angular components are nonzero for these BPS equations. In the last section~\ref{sec 6}, we give conclusions and remarks.

\section{BPS Lagrangian Method}
\label{sec 2}

At first let us write the ansatz for scalar and $U(1)$ gauge fields in the polar coordinates as follows
\begin{equation}
\label{ansatz}
 \phi=f(r) e^{in\theta}, \qquad \qquad \qquad \qquad A_\theta=a(r)-n,
\end{equation}
where $n=\pm1,\pm2,\dots$ is the winding number and we have set the gauge coupling to be unity for simplification.
Here we consider static configuration with temporal and radial gauge, $A_t=A_r=0$. Although the ansatz (\ref{ansatz}) depends on angle coordinate $\theta$, the effective Lagrangian $\mathcal{L}_{eff}$ will be eventually independent of $\theta$, and $n$ as well. As an example the (standard) kinetic terms are
\begin{equation}
\label{kinetic f}
 |D_\mu \phi|^2=-f'^2-\left(af \over r\right)^2,
\end{equation}
and
\begin{equation}
\label{kinetic a}
 B^2\equiv {1\over 2} F_{\mu\nu} F^{\mu\nu}= \left(a' \over r\right)^2,
\end{equation}
with $D_\mu\phi=\partial_\mu \phi+i A_\mu \phi$ and the signature of the metric is taken to be $(+,-,-)$. We use $'\equiv {\partial \over \partial r}$ unless there is an explicit argument in the function then it means taking derivative over the argument. In this case the effective Lagrangian will be a function $\mathcal{L}_{eff}(a',f',a,f;r)$.

In this case the BPS Lagrangian is simply written as~\cite{Atmaja:2015umo}
\begin{equation}
\label{Lagrangian BPS}
 L_{BPS}=\int dQ=\int d^2x \left({\partial Q\over \partial a}{a'\over 2\pi r}+{\partial Q\over \partial f}{f'\over 2\pi r}\right)=\int d^2x~ \mathcal{L}_{BPS},
\end{equation}
where $Q$ is called BPS energy function and it is assumed to be function of $a$ and $f$, but not of $r$ explicitly. Equating $\mathcal{L}_{eff}$ with $\CL_{BPS}$, $\CL_{eff}=\CL_{BPS}$, we may consider it as a quadratic equation for $a'$ or $f'$. One can try to split this equation into two quadratic equations for $a'$ and $f'$ separately. However in general splitting this equation may a bit tricky. We can consider it first as a quadratic equation for $a'$ and solve it using the quadratic formula that will give us two solutions, $a'_\pm$. We must set these two solutions to be equal, $a'_+=a'_-$, as a requirement that $\CL_{eff}$ can be rewritten in a complete square form. This can be done by setting the square-root term in the quadratic formula to be zero, which then becomes a quadratic equation of $f'$. Solving it and following the similar steps as before for $a'$, we obtain at the end an equation that does not contain $a'$ and $f'$. This last equation must be valid for all values of $r$ and so we can solve it order by order of power $r$. The BPS equations then are given by $a'=a'_\pm$ and $f'=f'_\pm$ with additional constraint equations that might come from solving the last equation. This last equation should be the same even if we consider the equation $\CL_{eff}=\CL_{BPS}$ as a quadratic equation for $f'$ first.

As an example let us consider the standard Maxwell-Higgs model with the following Lagrangian
\begin{equation}
  \CL=-{1\over 4}F_{\mu\nu}F^{\mu\nu}+\left|D_\mu\phi\right|^2-V(|\phi).
\end{equation}
Using the ansatz (\ref{ansatz}), its effective Lagrangian is given by
\begin{equation}
 \mathcal{L}_{eff}=-{1\over 2}\left(a' \over r\right)^2-\left(f'^2+\left(af \over r\right)^2\right)-V(f).
\end{equation}
Now, taking this effective Lagrangian to be equal to the BPS Lagrangian, $\CL_{BPS}$ in (\ref{Lagrangian BPS}), give us an equation
\begin{equation}
 -{1\over 2}\left(a' \over r\right)^2-\left(f'^2+\left(af \over r\right)^2\right)-V(f)={\partial Q\over \partial a}{a'\over r}+{\partial Q\over \partial f}{f'\over r},
\end{equation}
where we have rescaled $Q\to 2\pi Q$ for simplification.
Consider it as quadratic equation for $a'$ first, the two solutions are
\begin{equation}
 a'_\pm={-{\partial Q\over \partial a}~r \pm \sqrt{-2r^2f'^2-2r{\partial Q\over \partial f} f'+\left(\partial Q\over \partial a\right)^2 r^2-2r^2V-2a^2f^2}}.
\end{equation}
The two solutions will be equal if
\begin{equation}
 -2r^2f'^2-2r{\partial Q\over \partial f} f'+\left(\partial Q\over \partial a\right)^2 r^2-2r^2V-2a^2f^2=0,
\end{equation}
which is a quadratic equation for $f'$. Solutions to this equations are
\begin{equation}
 f'_\pm=-{1\over 2r}{\partial Q\over \partial f}\pm {1\over 2}\sqrt{r^2\left(-4 a^2 f^2 +\left(\partial Q\over \partial f\right)^2 + 2 r^2 \left(\left(\partial Q\over \partial a\right)^2 - 2 V\right)\right)}
\end{equation}
thus give us the last equation, upon equation these solutions,
\begin{equation}
 -4 a^2 f^2 ++\left(\partial Q\over \partial f\right)^2 + 2 r^2 \left(\left(\partial Q\over \partial a\right)^2 - 2 V\right)=0.
\end{equation}
Solving the last equations order by order of power $r$ yields two equations
\begin{eqnarray}
 {\partial Q\over \partial a}&=&\pm\sqrt{2V},\\
 {\partial Q\over \partial f}&=&\pm 2af,
\end{eqnarray}
which have solution $Q=\pm a\left(f^2-1\right)$ and $V={1\over 2}\left(f^2-1\right)^2$. The BPS equations are then given by
\begin{eqnarray}
 {a'\over r}&=&\pm\left(1-f^2\right),\\
 f'&=&\mp {af \over r}.
\end{eqnarray}

\section{Generalized Maxwell-Higgs Model}
\label{sec 3}

The generalized Maxwell-Higgs model described by the following Lagrangian~\cite{Bazeia:2012uc}
\begin{equation}
 \CL_{GenMH}=-{G(|\phi|) \over 4}F_{\mu\nu}F^{\mu\nu}+w(|\phi|)\left|D_\mu\phi\right|^2-V(|\phi),
\end{equation}
and the effective Lagrangian, using the ansatz (\ref{ansatz}), is given by
\begin{equation}
\label{Gen Lagrangian}
 \mathcal{L}_{GenMH}=-{G(f)\over 2}\left(a' \over r\right)^2-w(f)\left(f'^2+\left(af \over r\right)^2\right)-V(f),
\end{equation}
with $G(f),w(f)>0$ and $V(f)\geq0$.
Taking the BPS energy function to be $Q=2\pi F(f) A(a)$, let us write the BPS Lagrangian to be\footnote{This form of BPS Lagrangian was suggested in~\cite{Atmaja:2015lia}.}
\begin{equation}
\label{BPS Lagrangian}
 \CL_{BPS}=-F'(f) {A\over r} f'-A'(a) F(f) {a'\over r}-X_2(a,f,r) f'^2-X_0(a,f,r). 
\end{equation}
Here we have assumed that $Q$ does not depend explicitly on coordinate $r$ and is a separable function of fields $f$ and $a$ as in the previous case. The last two additional terms in the above BPS Lagrangian are necessary and they are related to each other by means of eliminating one of them will make the other to be set to zero. This BPS Lagrangian is different from the resulting BPS energy density of vortices, with non-zero pressure, computed in~\cite{Atmaja:2015umo}. We will show later using this BPS Lagrangian we can reproduce the BPS vortex, with non-zero internal pressure, as in~\cite{Atmaja:2015lia}.

Equating both Langrangians and collecting all terms that contains derivative of $f$, we then have
\begin{equation}
w(f)\left(f'^2+\left(af \over r\right)^2\right)=X_2(a,f,r) f'^2+F'(f) {A\over r} f'.
\end{equation}
Solution to this equation is
\begin{equation}
 (w-X_2)\left(f'\mp {af\over r}\sqrt{w\over w-X_2}\right)^2=0,
\end{equation}
with a constraint equation
\begin{equation}
 F'(f) A=\pm 2~a~f\sqrt{w(w-X_2)}.
\end{equation}
We may conclude from this constraint equation that $A=a$ and $X_2\equiv X_2(f)$. So, we obtain a BPS equation
\begin{equation}
\label{BE1}
f'=\pm {af\over r}\sqrt{w\over w-X_2}
\end{equation}
and the constraint equation
\begin{equation}
\label{CE1}
 F'(f)=\pm 2~f\sqrt{w(w-X_2)}.
\end{equation}
For the remaining terms, we have
\begin{equation}
 {G(f)\over 2}\left(a' \over r\right)^2+V(f)=A'(a) F(f) {a'\over r}+X_0(a,f,r).
\end{equation}
Solution to this equation is
\begin{equation}
 {G\over 2}\left({a'\over r}\mp \sqrt{2(V-X_0)\over G}\right)^2=0,
\end{equation}
with a constraint equation
\begin{equation}
\label{CE2}
 F=\pm \sqrt{2G(V-X_0)}.
\end{equation}
This constraint equation also implies $X_0\equiv X_0(f)$. So, we have another BPS equation
\begin{equation}
\label{BE2}
 {a'\over r}=\pm \sqrt{2(V-X_0)\over G}
\end{equation}
along with the other constraint equation (\ref{CE2}).

Now we still have two arbitrary functions $X_2$ and $X_0$ in the BPS Lagrangian (\ref{BPS Lagrangian}). Using the two constraint equations (\ref{CE1}) and (\ref{CE2}), we reduce the number of arbitrary functions by one such that
\begin{equation}
\label{CE3}
 {\partial \over \partial f}\sqrt{2G(V-X_0)}=2f\sqrt{w(w-X_2)}.
\end{equation}
Having BPS equations (\ref{BE1}) and (\ref{BE2}), we can rewrite the effective Lagragian (\ref{Gen Lagrangian}) to become
\begin{eqnarray}
\label{explicit eff Lagragian}
 \CL_{eff}&=&w\left(f'\mp {af\over r}\sqrt{w\over w-X_2}\right)^2-\left(af\over r\right)^2{wX_2 \over w-X_2}\pm 2w{a\over r}ff'\sqrt{w\over w-X_2}\nonumber\\
 &&+{G\over 2}\left({a'\over r}\mp \sqrt{2(V-X_0)\over G}\right)^2+X_0\pm {a'\over r}\sqrt{2G(V-X_0)}.
\end{eqnarray}
One can check that its Euler-Lagrange equation for $a$ is simply just the equation (\ref{CE3}) after substituting the BPS equations (\ref{BE1}) and (\ref{BE2}). Its Euler-Lagrange equation for $f$ is given by
\begin{eqnarray}
 \pm{\partial \over \partial r}\left(2waf\sqrt{w\over w-X_2}\right)&=&-{a^2\over r} {\partial\over \partial f}\left(wf^2X_2 \over w-X_2\right)\pm 2af' {\partial\over \partial f}\left(wf\sqrt{w\over w-X_2}\right)\nonumber\\
 &&+rX_0'(f)\pm{a'}{\partial\over \partial f}\sqrt{2G(V-X_0)}.
\end{eqnarray}
Using the BPS equation (\ref{BE2}) and the constraint equations (\ref{CE3}), we can simplify the Euler-Lagrange equation for $f$ to be
\begin{equation}
\label{constraint eqn}
 {a^2\over r^2} {\partial \over \partial f}\left(wf^2X_2 \over w-X_2\right)=X_0'(f)-2fX_2\sqrt{2w(V-X_0) \over G(w-X_2)}.
\end{equation}
The left hand side of equation, which contain function $a$ and coordinate $r$ explicitly, should be zero since all terms on the right hand side of equation are only functions of $f$. Another reason is because the equation (\ref{constraint eqn}) must be true for all values of $r$ and so we can solve it order by order in power of $r$ explicitly which then implies the left hand side of equation should be zero. This yields
\begin{equation}
\label{Sol X_2}
 {wf^2X_2 \over w-X_2}=C_0\qquad\longrightarrow\qquad X_2={w~C_0 \over f^2w+C_0},
\end{equation}
where $C_0$ is a constant. Let us write $X_0=V-{G\over 2}R^2$, with $R\equiv a'/r$, by using the BPS equation (\ref{BE2}). The equation (\ref{constraint eqn}) then can be written as
\begin{equation}
\label{add constraint}
 {\partial \over \partial f}\left(V-{G\over 2}R^2\right)=2w~C_0 {R\over f^2w+C_0},
\end{equation}
which is the same constraint equation as in~\cite{Atmaja:2015lia}. Notice that $C_0=0$ if $X_0$ is just a constant and this also implies $X_2=0$ by the right hand side of constraint equation (\ref{constraint eqn}). As shown in~\cite{Atmaja:2015lia}, finite energy condition requires that
\begin{equation}
 \int r \left(V-{G\over 2}\left({a'\over r}\right)^2\right) dr=\int r~X_0~dr=0
\end{equation}
The constant value of $X_0$ will be fixed to zero by the above finite energy condition. Therefore zero value of $X_0$ is related to zero value of $X_2$. Using the equation (\ref{CE3}), we can write $R$, or the BPS equation (\ref{BE2}), as
\begin{equation}
 R=\pm{1\over G}\left(2\int df {f^2 w^{3/2}\over \sqrt{f^2w+C_0}}+C_1\right),
\end{equation}
with $C_1$ is an integration constant.

\subsection{Euler-Lagrange equations of BPS Lagrangian}
Another way to obtain the additional constraint (\ref{add constraint}) is by deriving Euler-Lagrange equation of the BPS Lagrangian (\ref{BPS Lagrangian}). The first two terms of the BPS Lagrangian (\ref{BPS Lagrangian}) are
\begin{equation}
 \propto\int dr \left({\partial Q \over \partial a} a'+{\partial Q \over \partial f} f'\right)
\end{equation}
whose Euler-Lagrange equations are
\begin{eqnarray}
{d\over dr}\left({\partial Q \over \partial a}\right)={\partial^2 Q \over \partial a^2} a'+{\partial^2 Q \over \partial a\partial f}f' \\
{d\over dr}\left({\partial Q \over \partial f}\right)={\partial^2 Q \over \partial f^2} f'+{\partial^2 Q \over \partial a\partial f}a'.
\end{eqnarray}
The left hand side of both equations above is equal to its right hand side and so they are trivially zero. One can also argue that those two terms do not contribute to the Euler-Lagrange equations since they are boundary terms. The remaining two terms in the BPS Lagrangian (\ref{BPS Lagrangian}) are
\begin{equation}
 \propto\int dr~r\left(X_0(f)+X_2(f) f'^2\right)
\end{equation}
whose Euler-Lagrange equation is\footnote{There is only Euler-Lagrange equation for $f$, since those terms only depend on field $f$.}
\begin{equation}
 2X_2{d\over dr}\left(r f'\right)+r{\partial X_2\over \partial f}f'^2=r{\partial X_0 \over \partial f}. \label{EL BPS Lagrangain}
\end{equation}
In the BPS limit, in which the BPS equations (\ref{BE1}) and (\ref{BE2}) are satisfied, this equation is equal to the equation (\ref{constraint eqn}) and hence later gives us the constraint equation (\ref{add constraint}). Therefore it is not necessary to write down the effective Lagrangian in a complete square-forms (\ref{explicit eff Lagragian}), but instead we could use the BPS Lagrangian (\ref{BPS Lagrangian}) and derive its Euler-Lagrange equations as additional constraint equations.

\subsection{General BPS Lagrangian}
One may ask what happens if we add a term which is proportional to $a'^2$ into the BPS Lagrangian (\ref{BPS Lagrangian}). Let us now consider a more general BPS Lagrangian as follows
\begin{equation}
\label{BPS Lagrangian general}
 \CL_{BPS}=-F'(f) {A\over r} f'-A'(a) F(f) {a'\over r}-X_0(a,f,r)-X_1(a,f,r) a'^2-X_2(a,f,r) f'^2. 
\end{equation}
Equating it with the effective Lagrangian (\ref{Gen Lagrangian}), $\CL_{GenMH}=\CL_{BPS}$, we may consider it first as a quadratic equation for $a'$ which has solutions
\begin{eqnarray}
 a'_\pm&=&{A'(a)F~r \pm \sqrt{S_a}  \over G- 2r^2X_1}, \label{BPS a}\\
 S_a&=&A'(a)^2F^2r^2-2\left(G- 2r^2X_1\right)\left(-A~F'(f) f' r+a^2f^2w+r^2\left(V-X_0+f'^2\left(w-X_2\right)\right)\right).\nonumber\\
\end{eqnarray}
The two solutions will be equal if $S_a=0$. The later equation can be considered as a quadratic equation for $f'$ which has solutions
\begin{eqnarray}
 f'_\pm&=&{A~F'(f)r\left(G- 2r^2X_1\right) \pm \sqrt{S_f}  \over 2r^2\left(G- 2r^2X_1\right)\left(w-X_2\right)}, \label{BPS f}\\
 S_f&=&r^2\left(G- 2r^2X_1\right)\left(2\left(w-X_2\right)\left(A'(a)^2F^2r^2-2\left(G- 2r^2X_1\right)\left(a^2f^2w+r^2\left(V-X_0\right)\right)\right)\right.\nonumber\\
 &&\left. +A^2F'(f)^2\left(G- 2r^2X_1\right)\right).
\end{eqnarray}
Similarly two solutions will be equal if $S_f=0$. Here $S_f$ does not contain any first-derivative of the effective fields, $a'$ and $f'$, and so we can solve it algebraically. But at first notice that in order for BPS equations (\ref{BPS a}) and (\ref{BPS f}) to be well-defined we must have $\left(G- 2r^2X_1\right)\neq 0$ and $\left(w-X_2\right)\neq0$. The equation $S_f=0$ can be expanded in terms of explicit power of $r$ as follows
\begin{eqnarray}
\label{Gen Constraint}
 0&=&{Q'(f)^2\over 4\pi^2}G-4a^2f^2G~w\left(w-X_2\right)+2\left(\left(w-X_2\right)\left({Q'(a)^2\over 4\pi^2}+4a^2f^2w~X_1-2G\left(V-X_0\right)\right)\right.\nonumber\\
 &&\left.-{Q'(f)^2\over 4\pi^2}X_1\right)r^2+8X_1\left(V-X_0\right)\left(w-X_2\right)r^4,
\end{eqnarray}
with $Q=2\pi~A(a)~F(f)$. The above equation must be true for all values of $r$. So, we could try to solve it order by order in explicit power of $r$. There we have three equations in terms of explicit power of $r^0, r^2,$ and $r^4$. Besides these equations, we have at least additional two constraint equations comming from Euler-Lagrange equations of the BPS Lagrangian (\ref{BPS Lagrangian general}) as shown in the previous section\footnote{The number constraint equations could be larger than two, since each Euler-Lagrange equations may have explicit power of $r$ and hence must be expanded an solved for each explicit power of $r$ as well.}. On the other hand we have four arbitrary functions $Q, X_0, X_1,$ and $X_2$ in the BPS Lagrangian (\ref{BPS Lagrangian general}). So naively there are four arbitrary functions and five constraint equations which means we have overdetermined set of constraint equations. This is even worst if we consider the functions $X_0, X_1,$ and $X_2$ to depend explicitly on $r$, which increases the number of constraint equations. To simplify the problem, we assume that the functions $X_0, X_1,$ and $X_2$ do not depend explicitly on $r$.

Our task now is to make the number of arbitrary functions and of constraint equations to be equal. The number of Euler-Lagrange equations of the BPS Lagrangian will be reduced to one if the functions $X_0, X_1,$ and $X_2$ only depend on field $a$ or $f$. In this case, we further assume that those functions only depend on field $f$ and by this the number of constraint equations is now equal to the number of arbitrary functions\footnote{We found that this is a good choice. Dependency on field $a$ in one of the unknown functions $X_0, X_1,$ and $X_2$ will make the analysis becomming more complicated.}. Then from the $r^4$-order equation of (\ref{Gen Constraint}), we can take $V=X_0$ while the $r^0$-order equation yields
\begin{equation}
\label{r^0 equation}
 {Q'(f)^2\over 4\pi^2}=4a^2f^2w\left(w-X_2\right)
\end{equation}
as such $Q\propto a$ and $w>X_2$. This equation is equal to the constraint equation (\ref{CE1}). However, substituting these solutions to the $r^2$-order equation forces us to take $w=X_2$ which contradicts with previous condition that the BPS equation (\ref{BPS f}) must be well-defined. Rather then taking $V=X_0$ in the $r^4$-order equation, we could take $X_1=0$ which also means we reduce the number of arbitrary functions by one. Fortunately, this also reduces the number of constraint equations by one, or more precisely there is no $r^4$-order equation in (\ref{Gen Constraint}). The $r^2$-order equation now becomes
\begin{equation}
{Q'(a)^2\over 4\pi^2}-2G\left(V-X_0\right)=0
\end{equation}
which does not depend on field $a$ explicitly and is equal to the constraint equation (\ref{CE2}). Therefore we should take $X_1=0$ in (\ref{BPS Lagrangian general}), which is equal to the BPS Lagrangian (\ref{BPS Lagrangian}), and we get the same results as before. Notice that the number of constraint equations is actually five not four, since the constraint equations comming from the Euler-Lagrange equation (\ref{EL BPS Lagrangain}) are two instead of one. This however is not a problem because the excessive contraint equation will reduce the number of arbitrary functions ($w$,$G$,$V$) in the effective Lagrangian (\ref{Gen Lagrangian}).

\section{Generalized Born-Infeld-Higgs Model}
\label{sec 4}

Let us now extend the generalized Maxwell-Higgs model (\ref{Gen Lagrangian}) into the Born-Infeld type of action~\cite{Born:1934gh,Shiraishi:1990zi} in which the effective action is given by~\cite{Casana:2015bea}
\begin{equation}
 \CL_{GenBIH}=-b^2\left(\sqrt{1+{G(|\phi|) \over 2 b^2}F_{\mu\nu}F^{\mu\nu}}-1\right)+w(|\phi|)\left|D_\mu\phi\right|^2-V(|\phi),
\end{equation}
with $G,w>0$, $V\geq0$ are arbitrary functions of $f$, and $b$ is the Born-Infeld parameter.
Using the ansatz (\ref{ansatz}), we obtain an effective Lagrangian\begin{equation}
\label{Lagrangian GenBIH}
 \CL_{GenBIH}=-b^2\left(\sqrt{1+{1\over b^2}G(f){a'^2 \over r^2}}-1\right)-w(f)\left(f'^2+{a^2f^2 \over r^2}\right)-V(f).
\end{equation}
We will be using the following general BPS Lagrangian
\begin{equation}
\label{BPS Lagrangian GenBIH}
 \CL_{BPS}=-{Q'(f)\over r} f'-{Q'(a)\over r}a'-X_0(a,f,r)-X_1(a,f,r) a'-X_2(a,f,r) f'^2, 
\end{equation}
with $Q\equiv Q(a,f)$. Unlike the BPS Lagrangian (\ref{BPS Lagrangian general}), the fourth term on right hand side is proportional to $a'$ instead of $a'^2$. The reason is because the equation $\CL_{GenBIH}=\CL_{BPS}$ will look like a quartic equation for $a'$, which could lead to a more complicated analysis.

Equating both Lagrangians (\ref{Lagrangian GenBIH}) and (\ref{BPS Lagrangian GenBIH}), and considering it first as quadratic equations for $f'$, we obtain solutions
\begin{eqnarray}
 f'_\pm&=&{Q'(f)\pm\sqrt{S_f} \over 2 r\left(w-X_2\right)},\label{BPS f-2}\\
 S_f&=& Q'(f)^2+4\left(w-X_2\right)\left(a'r\left(Q'(a)+rX_1\right)-r^2\left(V-X_0\right)-a^2f^2w\right.\nonumber\\
 &&\left.-b^2 r^2\left(\sqrt{1+{1\over b^2}G{a'^2 \over r^2}}-1\right)\right).
\end{eqnarray}
The two solutions will be equal if $S_f=0$ which is later considered as quadratic equation for $a'$ and has solutions
\begin{eqnarray}
 a'_\pm&=&\frac{(Q'(a)+r X_1) \left(4 (w-X_2) \left(r^2 \left(V-X_0-b^2\right)+a^2 f^2 w\right)-Q'(f)^2\right)\pm\sqrt{S_a}}{4 r
   (w-X_2) \left((Q'(a)+r X_1)^2-b^2 G\right)},\label{BPS a-2}\\
   S_a&=&b^2 r^2 (w-X-2)^2 \left(G \left(Q'(f)^2-4 (w-X-2) \left(a^2 f^2 w+r^2 (V-X_0)\right)\right)\right.\nonumber\\
   &&\left.\left(4 (w-X-2) \left(r^2 \left(2
   b^2-V+X_0\right)-a^2 f^2 w\right)+Q'(f)^2\right)\right.\nonumber\\
   &&\left.+16 b^2 r^4 (w-X_2)^2 (Q'(a)+r X_1)^2\right).
\end{eqnarray}
The solutions will be equal if $S_a=0$ and, as previously disussed, we solve it order by order in explicit power of $r$ as such we obtain several constraint equations
\begin{eqnarray}
 r^2 &:& b^2 G (w-X_2)^2 \left(4 a^2 f^2 w (X_2-w)+Q'(f)^2\right)^2=0,\label{r2}\\
 r^4 &:& 8 b^2 G (b^2 - V + X_0) (w - X_2)^3 (Q'(f)^2 + 4 a^2 f^2 w (-w + X_2))=0,\label{r4}\\
 r^6 &:& 16 b^2 (G (V - X_0)^2 + b^2 (Q'(a)^2 + 2 G (-V + X_0))) (w - X_2)^4=0,\label{r6}\\
 r^7 &:& 32 b^4 Q'(a) X_1 (w - X_2)^4=0,\label{r7}\\
 r^8 &:& 16 b^4 X_1^2 (w - X_2)^4=0.\label{r8}
\end{eqnarray}
There are at least additional two constraint equations, comming from the Euler-Lagrange equations of BPS Lagrangian (\ref{BPS Lagrangian GenBIH}), and four arbitrary functions in the BPS Lagrangian (\ref{BPS Lagrangian GenBIH}) in which we have taken $X_0, X_1,$ and $X_2$ are explicitly independent of $r$. From equation (\ref{r8}), we conclude that $X_1=0$, since $w=X_2$ is prohibited by the BPS equation (\ref{BPS f-2}) and so again like in the previous section the BPS Lagrangian (\ref{BPS Lagrangian GenBIH}) is equal to the BPS Lagrangian (\ref{BPS Lagrangian}). This will reduce the number of arbitrary functions by one and the number of constraint equations by two, which are equations (\ref{r7}) and (\ref{r8}). As mentioned previously we can further reduce the number of constraint equations from Euler-Lagrange equations of BPS Lagrangian (\ref{BPS Lagrangian GenBIH}) by assuming $X_0$ and $X_2$ are only functions of $f$. Furthermore, the equation (\ref{r2}) is solved by
\begin{equation}
 Q'(f)^2=4 a^2 f^2 w (w-X_2).\label{Solution GenBIH-1}
\end{equation}
which also solves the equation (\ref{r4}). Therefore those two consraint equations are actually redundant and thus reduce the number of constraint equations by one. At this state the number of arbitrary functions is three and the number of constraint equations is four, since the number of constraint equations from the Euler-Lagrange equation (\ref{EL BPS Lagrangain}) are two. As mentioned previously this will not be an issue because there are actually another arbitrary functions ($w$,$G$,$V$) in the effective Lagragian (\ref{Lagrangian GenBIH}).

The equation (\ref{Solution GenBIH-1}) implies $Q\propto a$ and then the equation (\ref{r6}) give us
\begin{equation}
Q^2={a^2 \over b^2}G \left(V-X_0\right)\left(2b^2-\left(V-X_0\right)\right),
\end{equation}
with $2b^2>V-X_0>0$, and a constraint equation
\begin{equation}\label{CE4}
  {\partial\over\partial f}\sqrt{G \left(V-X_0\right)\left(2b^2-\left(V-X_0\right)\right)}=2f\sqrt{b^2w\left(w-X_2\right)}.
\end{equation}
Using the previous results, the BPS equations now become
\begin{eqnarray}
 {a'\over r}&=&\pm\frac{\sqrt{b^2 G\left(V-X_0\right)\left(2b^2-V+X_0\right)}}{G\left(b^2-V+X_0\right)},\label{BPS BIH a}\\
 f'&=&\pm{af \over r}\sqrt{w \over w-X_2}.\label{BPS BIH f}
\end{eqnarray}
Besides these BPS equations, we also still have constraint equations (\ref{CE4}) and (\ref{EL BPS Lagrangain}). Substituting the above BPS equations into the constraint equation (\ref{EL BPS Lagrangain}) and solving it order by order in explicit power of $r$ yield
\begin{eqnarray}
 r^{-1} &:& {\partial \over \partial f}\left({f^2X_2w \over w-X_2}\right)=0, \label{constraint 1}\\
 r^1 &:& 2X_2f\sqrt{w \over w-X_2}=\frac{G\left(b^2-V+X_0\right)}{\sqrt{b^2 G\left(V-X_0\right)\left(2b^2-V+X_0\right)}} {\partial X_0 \over \partial f}. \label{constraint 2}
\end{eqnarray}
The equation (\ref{constraint 1}) has solution given by (\ref{Sol X_2}) which is similar to the case of generalized Maxwell-Higgs model. Substituting this solution, along with the BPS equations (\ref{BPS BIH a}) and (\ref{BPS BIH f}), into equation (\ref{constraint 2}) gives us
\begin{equation}
X_0'(r)=C_0 {(a^2)'\over r^2}.\label{constraint X0}
\end{equation}
This equation is also valid for the case of generalized Maxwell-Higgs model. The remaining constraint equation (\ref{CE4}) will fix one of arbitrary functions in the effective Lagrangian (\ref{Lagrangian GenBIH}).

\section{Energy-Momentum Tensor}
\label{sec 5}

The energy-momentum tensor for generalized Born-Infeld-Higgs model is given by~\cite{Casana:2015bea}
\begin{equation}
 T_{\mu\nu}=-{G F^\alpha_{\ \mu} F_{\alpha\nu}\over \sqrt{1+{1\over2 b^2}G F_{\mu\nu}F^{\mu\nu}}}+w \left(\overline{D_\mu\phi}D_\nu\phi+D_\mu\phi \overline{D_\nu\phi}\right)-\eta_{\mu\nu}\CL.
\end{equation}
Substituting the ansatz (\ref{ansatz}) and writing in polar coordinate, the non-zero components of the energy-momentum tensor are
\begin{eqnarray}
 T_{tt}&=&b^2\left(\sqrt{1+{G\over b^2}{a'^2\over r^2}}-1\right)+w\left(f'^2+{a^2f^2 \over r^2}\right)+V,\\
 T_{rr}&=&{b^2\over\sqrt{1+{G\over b^2}{a'^2\over r^2}} }\left(\sqrt{1+{G\over b^2}{a'^2\over r^2}}-1\right)+w \left(f'(r)^2-\frac{a(r)^2
   f(r)^2}{r^2}\right)-V,\label{Tradial}\\
 T_{\theta\theta}&=&{r^2b^2\over\sqrt{1+{G\over b^2}{a'^2\over r^2}} }\left(\sqrt{1+{G\over b^2}{a'^2\over r^2}}-1\right)-r^2w \left(f'(r)^2-\frac{a(r)^2
   f(r)^2}{r^2}\right)-r^2V.\label{Tangle}
\end{eqnarray}
Using the BPS equations, (\ref{BPS BIH a}) and (\ref{BPS BIH f}), and solution for $X_2$, (\ref{Sol X_2}), they are simplified further to
\begin{eqnarray}
 T_{tt}&=&{b^4\over b^2-V+X_0}-\left(b^2-V\right)+\left(2f^2w+C_0\right){a^2 \over r^2},\label{energy dens}\\
 T_{rr}&=&{a^2 \over r^2}C_0-X_0,\label{radial pressure}\\
 T_{\theta\theta}&=&-r^2\left({a^2 \over r^2}C_0+X_0\right).\label{angular pressure}
\end{eqnarray}
We can compute conservation of energy-momentum tensor, $\nabla_\mu T^{\mu\nu}=0$, in the polar coordinates. This would give additional constraint equations to our disposal. Suprisingly, the only non-trivial equation comes out from the conservation of energy-momentum tensor is equal to the constraint equation (\ref{constraint X0}), and hence no additional constraint equations produced.

\subsection{Finiteness of energy}
Let us consider one-parameter family of solutions by rescaling the space with a constant $0<\lambda<\infty$, $x\to {x\over\lambda}$, and now total static energy is given by a function of $\lambda$,
\begin{equation}
 E(\lambda)=\int d^2x\left[{b^2\over \lambda^2}\left(\sqrt{1+{\lambda^4 G \over 2 b^2}F_{ij}^2}-1\right)+w\left|D_i\phi\right|^2+{V\over\lambda^2}\right].
\end{equation}
Due to Derrick's theorem~\cite{Derrick:1964ww}, the total static energy can be finite if its extremum over $\lambda$ is given by a finite and positive value of $\lambda$. The extremum of total static energy is given by the following equation
\begin{equation}
 {dE \over d\lambda}=\int d^2x\left[{2b^2\over\lambda^3\sqrt{1+{\lambda^4G\over 2b^2}F_{ij}^2}}\left(\sqrt{1+{\lambda^4 G \over 2 b^2}F_{ij}^2}-1\right)-{2\over\lambda^3}V\right]=0.
\end{equation}
Now, it is rather difficult to prove that solution to this equation is at positive and finite $\lambda$ because there is $\lambda$ inside the square root. However we can prove it by considering the equation above to be satisfied locally such that
\begin{equation}
 {b^2\over\sqrt{1+{\lambda^4G\over 2b^2}F_{ij}^2}}\left(\sqrt{1+{\lambda^4 G \over 2 b^2}F_{ij}^2}-1\right)=V,
\end{equation}
which then gives us
\begin{equation}
 \lambda^4=2b^2 {V(2b^2-V) \over \left(b^2-V\right)^2 F_{ij}^2}>0,
\end{equation}
with $0<V< 2b^2$, $V\neq b^2$, and $F_{ij}^2>0$. So there is positive and finite value of $\lambda$ that can extrimize the static energy density, which also extrimize the total energy. Setting $\lambda=1$, we obtain a virial identity
\begin{equation}
 {b^2\over\sqrt{1+{G\over 2b^2}F_{ij}^2}}\left(\sqrt{1+{G \over 2 b^2}F_{ij}^2}-1\right)=V.\label{virial identity}
\end{equation}
Using the ansatz (\ref{ansatz}), it becomes
\begin{equation}
 {G\over b^2}\left(a'\over r\right)^2={V(2b^2-V)\over \left(b^2-V\right)^2},
\end{equation}
which is equal to the BPS equation (\ref{BPS BIH a}) if we shift $V\to V+X_0$. However, we could consider the virial identity globally by means of taking its integral over the whole space as such
\begin{equation}
 \int d^2x\left[ {b^2\over\sqrt{1+{G\over 2b^2}F_{ij}^2}}\left(\sqrt{1+{G \over 2 b^2}F_{ij}^2}-1\right)-V\right]=\int d^2x~ Y(f)=0.
\end{equation}
Substituting the ansatz (\ref{ansatz}) and the BPS equation (\ref{BPS BIH a}) into the above integran, we find
\begin{equation}
 \int dr~ r~ X_0(f(r))=0.\label{int X_0}
\end{equation}
Therefore solutions to the BPS equations (\ref{BPS BIH a}) and (\ref{BPS BIH f}) could give a finite total energy and non-zero internal pressure if $X_0$ is not constant and its intergral over whole space is zero. If $X_0$ is constant then it will be fixed to zero by the finite energy condition (\ref{int X_0}) and also it implies $C_0=0$ by the constraint equation (\ref{constraint X0}). This also means the pressure densities in radial and angular directions, $T_{rr}$ and $T_{\theta\theta}$ respectively, are zero as can be seen explicitly in the formulas (\ref{radial pressure}) and (\ref{angular pressure}).

\section{Numerical Analysis}

Finding analytical solutions of the BPS equations (\ref{BPS BIH a}) and (\ref{BPS BIH f}), together with the constraint equations (\ref{CE4}), (\ref{constraint X0}), and (\ref{int X_0}), is very difficult. Therefore, we will look for numerical solutions instead. The boundary values of effective fields $a$ and $f$ are
\begin{eqnarray}
 a(0)&=&n,\qquad\qquad\qquad a(\infty)= 0,\\
 f(0)&=&0,\qquad\qquad\qquad f(\infty)= 1.
\end{eqnarray}
It was suggested in~\cite{Atmaja:2015lia} that a type of functions satisfies the finite energy condition (\ref{int X_0}) is provided by the Laguerre functions $L_m$ as such
\begin{equation}
 X_0(r)= C_x~e^{-r} L_m(r),\qquad\qquad\qquad L_m(r)=e^{r}{d^m \over dr^m}\left(r^m e^{-r}\right),
\end{equation}
with $C_x\neq0$ is an arbitrary constant and $m=2,3,4,\ldots$. Let us check if these functions satisfy the constraint equation (\ref{constraint X0}) near the origin $r=0$. Near the origin we can expand the solution for $a$ as follows
\begin{equation}
 a(r)=n+a_0~ r^{\at_0}+\cdots,
\end{equation}
with $\at_0>0$ and $a_0\neq0$ are constant and finite. Substituting $X_0$ and near origin expansion of $a$ into the constraint equation (\ref{constraint X0}), the leading order implies $\at_0=3$ and $a_0=-{(m+1) C_x\over 6~C_0 n}$. We can continue further to find the next order solution of $a$ by writing near origin expansion of $a$ to be
\begin{equation}
 a(r)=n-{(m+1) C_x\over 6~C_0 n}r^3 + a_1~ r^{3+\at_1}+\cdots,
\end{equation}
with $a_1\neq0$ and $\at_1>0$ are constant and finite. The first two leading order of $C_0(a^2)'-r^2 X_0'(r)$ are
\begin{equation}
 (m-2)C_x~ r^2 +(6+2 \at_1)a_1 C_0 n~ r^{2+\at_1}+\cdots.
\end{equation}
The leading order will be zero only for $m=2$. However the next leading order can never be set to zero which means the Laguerre functions do not satisfy the costraint equation (\ref{constraint X0}) near the origin. Thus the Laguerre functions are unfortunately not good solutions for $X_0$.

\subsection{Near origin expansions}

Now let us try to find possible near origin expansion of fields $a,f,w,G,$ and $S\equiv V-X_0$, which take the forms of
\begin{eqnarray}
\label{couplings expansion}
 w&=&w_0f(r)^{\wt_0}+\cdots,\qquad G=g_0f(r)^{\gt_0}+\cdots,\qquad S=s_0 f(r)^{\st_0}+\cdots,\\ 
 \label{fields expansion}
 a(r)&=&n+a_0 r^{\at_0}+\cdots,\qquad\qquad f(r)=f_0r^{\ft_0}+\cdots, 
\end{eqnarray}
with $\at_0>0$; $\ft_0>0$; $a_0\neq0$; $f_0\neq0$; $w_0\neq0$; $g_0\neq0$; $s_0\neq0; \wt_0,\gt_0,$ and $\st_0$ are constant and finite. If any of the coefficients of expansion, without ``tilde'', is zero then it means the corresponding function is exact. Substituting these near origin expansions into BPS equation (\ref{BPS BIH f}) the possible leading order terms are given by
\begin{equation}
 -C_0 n^2+(\ft_0^2-n^2)w_0 f_0^{2+\wt_0} r^{(2+\wt_0)\ft_0}+\cdots.
\end{equation}
There are two possibilities for vanishing leading order:
\begin{enumerate}
 \item $\wt_0=-2\longrightarrow \ft_0=n\sqrt{1+{C_0\over w_0}}$ with $C_0+w_0\neq0$.
 \item $\wt_0<-2\longrightarrow \ft_0=n$.
\end{enumerate}
Here we have assumed $n=1,2,3,\ldots$. First, we consider the possibility with $\wt_0=-2$ as such the expansions (\ref{couplings expansion}) can be written as
\begin{equation}
 w=w_0~f(r)^{-2}+w_1~f(r)^{-2+\wt_1}+\cdots,\qquad G=g_0~f(r)^{\gt_0}+\cdots,\qquad S=s_0~ f(r)^{\st_0}+\cdots,
\end{equation}
with $w_1\neq0$ and $\wt_1>0$ are constant and finite. Substituting these expansions into the constraint equation (\ref{CE4}), its possible leading order terms are
\begin{eqnarray}
 &\propto& -32 b^4 g_0 s_0 w_0^3+16 b^2 g_0 s_0^2
   w_0^3 f(r)^{\st_0}+4 b^4 g_0^2 s_0^2 (C_0+w_0) (\gt_0+\st_0)^2
   f(r)^{\gt_0+\st_0}\nonumber\\
   &&-4 b^2 g_0^2 s_0^3
   (C_0+w_0) (\gt_0+\st_0) (\gt_0+2\st_0) f(r)^{\gt_0+2\st_0}+g_0^2 s_0^4 (C_0+w_0) (\gt_0+2\st_0)^2
   f(r)^{\gt_0+3\st_0}+\cdots.\nonumber\\
\end{eqnarray}
The leading term is valid if only if $\st_0=0$ and $s_0=2b^2$. Taking $s_0=2b^2$, we rewrite the expansions (\ref{couplings expansion}) as
\begin{equation}
 w=w_0f(r)^{-2}+w_1f(r)^{-2+\wt_1}+\cdots,\qquad G=g_0f(r)^{\gt_0}+\cdots,\qquad S=2b^2+s_1 f(r)^{\st_1}+\cdots,
\end{equation}
with $s_1\neq0$; $\st_1>0$ are constant and finite. Substituting back these expansion into the constraint equation (\ref{CE4}) gives us possible leading terms
\begin{equation}
 \propto 32 b^4 g_0 s_1 w_0^3+4 b^4 g_0^2 s_1^2 (C_0+w_0) (\gt_0+\st_1)^2
   f(r)^{\gt_0+\st_1}+\cdots.
\end{equation}
The only possible way for these leading terms to be zero is when $s_1=0$ which implies a constant function $S=2b^2$, but this is not allowed by the BPS equation (\ref{BPS BIH a}) since the effective field $a$ will be trivial.

Now our only option is $\wt_0<-2$ in which, using the expansions (\ref{couplings expansion}), the possible leading terms in the constraint equation (\ref{CE4}) are
\begin{eqnarray}
 &\propto& b^4 g_0^2 s_0^2 w_0 (\gt_0+\st_0)^2 f(r)^{\gt_0+\st_0}-4 b^2 g_0^2 s_0^3
   w_0 (\gt_0+\st_0) (\gt_0+2\st_0) f(r)^{\gt_0+2\st_0}+g_0^2 s_0^4 w_0 (\gt_0+2\st_0)^2 f(r)^{\gt_0+3\st_0}\nonumber\\
   &&-32 b^4
   g_0 s_0 w_0^3 f(r)^{2\wt_0+4}+16 b^2 g_0 s_0^2 w_0^3
   f(r)^{\st_0+2\wt_0+4}+\cdots.
\end{eqnarray}
In this case there are few possible values of $\st_0$ and $\gt_0$:
\begin{itemize}
 \item $\st_0<0 \longrightarrow \gt_0=2(2+\wt_0-\st_0)$ and $g_0=-{4b^2 w_0^2 \over s_0^2(2+\wt_0)^2}~\forall \gt_0$.
 \item $\st_0=0 \longrightarrow s_0=2b^2 ~\forall \gt_0$ except for $\gt_0=4+2\wt_0$ in which $g_0=-{4b^2 w_0^2 \over s_0(s_0-2b^2)(2+\wt_0)^2}$.
 \item $\st_0>0 \longrightarrow \gt_0+\st_0=2(2+\wt_0)$ and $g_0={2 w_0^2 \over (2+\wt_0)^2 s_0}$ which is only for $\gt_0<0$.
\end{itemize}

\subsubsection{case of $\st_0<0$}
Writing the expansions (\ref{couplings expansion}) as
\begin{eqnarray}
  w&=&w_0 f(r)^{\wt_0}+\cdots,\qquad G=-{4b^2 w_0^2 \over s_0^2 (2+\wt_0)^2}f(r)^{2(2+\wt_0-\st_0)}+g_1 f(r)^{2(2+\wt_0-\st_0)+\gt_1}+\cdots,\nonumber\\
  S&=&s_0 f(r)^{\st_0}+\cdots,
\end{eqnarray}
with $g_1\neq0 $ and $\gt_1>0$ are constant and finite. The possible leading expansion of the constraint equation (\ref{CE4}) are
\begin{equation}
 \propto -\frac{128 b^6 w_0^5 (2+\wt_0-\st_0) f(r)^{2+\wt_0-\st_0}}{s_0
   (\wt_0+2)^3}+\frac{64 b^4 C_0 w_0^4}{(\wt_0+2)^2}-\frac{16 b^2 g_1
   s_0^2 w_0^3 (\gt_1+\wt_0+2) f(r)^{\gt_1+\wt_0+2}}{\wt_0+2}+\cdots.
\end{equation}
The only possiblity for the leading terms to be zero is if $2+\wt_0-\st_0<0$ and $\gt_1=-\st_0$ with $g_1=-{8 b^4 w_0^2 \over (2+\wt_0)^2 s_0^3}$. Substituting these constants into expansion of $G$ and repeating the above steps will result in expansion of $G$ as follows:
\begin{equation}
 G=-{4b^2 w_0^2 \over s_0^2 (2+\wt_0)^2}f(r)^{2(2+\wt_0-\st_0)}\sum^{m_{max}}_{m=0}\left({2 b^2 \over s_0} f(r)^{-\st_0}\right)^m,
\end{equation}
where for each $m$-th order of the expansion $2+\wt_0-m~\st_0<0$. Since $\st_0<0$, there is highest order of expansion which is given by $m_{max}=\left\lceil{2+\wt_0\over \st_0}\right\rceil-1$, with $\left\lceil{\dots}\right\rceil $ is the ceiling function where the number inside it is rounded up to the nearest integer, in which the expansion of $G$ above is truncated at $m=m_{max}$, and hence is an exact expansion in this case. We could write the expansion of $S$ instead by rewriting the expansion (\ref{couplings expansion}) as
\begin{eqnarray}
  w&=&w_0 f(r)^{\wt_0}+\cdots,\qquad G=-g_0 f(r)^{\gt_0}+\cdots,\qquad\nonumber\\
  S&=&\pm{2~w_0 \over (2 + \wt_0)}\sqrt{b^2\over g_0} f(r)^{2+\wt_0-{\gt_0\over 2}}+s_1 f(r)^{2+\wt_0-{\gt_0\over 2}+\st_1}+\cdots,
\end{eqnarray}
with $s_1\neq 0$; $\st_1>0$; $g_0>0$; and $2+\wt_0-{\gt_0\over 2}<0$. Here we have replaced $g_0\to-g_0$ to simplify the notation. Repeating the same steps as in the expansion of $G$ above, we obtain
\begin{eqnarray}
 S&=&\pm{2~w_0 \over (2 + \wt_0)}\sqrt{b^2\over g_0} f(r)^{2+\wt_0-{\gt_0\over 2}}+b^2\pm{b^2 \sqrt{b^2g_0} (2 + \wt_0)\over 4 w_0}f(r)^{-2-\wt_0 +{\gt_0\over 2}}\nonumber\\
 && \mp{b^4 g_0 \sqrt{b^2g_0} (2 + \wt_0)^3\over 4^3 w_0^3} f(r)^{3(-2-\wt_0 +{\gt_0\over 2})} \pm 
 {2~b^6 g_0^2\sqrt{b^2g_0} (2 + \wt_0)^5\over 4^5 w_0^5} f(r)^{5(-2-\wt_0 +{\gt_0\over 2})} \nonumber\\
 &&\mp{5~ b^8 g_0^3\sqrt{b^2g_0} (2 + \wt_0)^7\over 4^7 w_0^7} f(r)^{7(-2-\wt_0 +{\gt_0\over 2})} \pm 
 {14~ b^{10} g_0^4\sqrt{b^2g_0} (2 + \wt_0)^9\over 4^9 w_0^9} f(r)^{9(-2-\wt_0 +{\gt_0\over 2})}+\cdots,\nonumber\\
\end{eqnarray}
The first term on the right hand side of the expansion is valid if $4+2\wt_0-\gt_0<0$ while the second term is if $\gt_0<0$. The third and fourth terms are valid if $\gt_0-(2+\wt_0)<0$ and $2\gt_0-3(2+\wt_0)< 0$ respectively. The fifth, sixth, and seventh terms are valid if $3\gt_0-5(2+\wt_0)<0$, $4\gt_0-7(2+\wt_0)<0$, and $5\gt_0-9(2+\wt_0)<0$ respectively. Therefore for the $p$-th term is valid if $(p-2)\gt_0-(2p-5)(2+\wt_0)<0$ with $p\geq 3$. Since $2+\wt_0<0$ and $\gt_0<0$, there is $p_{max}$ terms in the expansion of $S$ at which $p_{max}=\left\lceil{2\gt_0-5(2+\wt_0)\over \gt_0-2(2+\wt_0)}\right\rceil-1$, with $2(2+\wt_0)<\gt_0<2+\wt_0$, and so the expansion of $S$ is exact for fixed $\gt_0$ and $\wt_0$. The expansion of $S$ above might have a simple expression given by\footnote{We have verified this expansion up to $m=4$. However we do not have general formula to prove it for higher order.}
\begin{eqnarray}
 S&=&\pm{2~w_0 \over (2 + \wt_0)}\sqrt{b^2\over g_0} f(r)^{2+\wt_0-{\gt_0\over 2}}+b^2\nonumber\\
 &&\pm{b^2 \sqrt{b^2g_0} (2 + \wt_0)\over 4 w_0}f(r)^{-2-\wt_0 +{\gt_0\over 2}}\sum^{m_{max}}_{m=0}\textbf{C}_m\left(-{b^2 g_0(2+\wt_0)^2 \over 16~w_0^2}\right)^m f(r)^{2m\left(-2-\wt_0 +{\gt_0\over 2}\right)},\nonumber\\
\end{eqnarray}
where $\textbf{C}_m={(2m)!\over (m+1)!m!}$ is the Catalan numbers and $m_{max}=\left\lceil{2+\wt_0-\gt_0\over \gt_0-2(2+\wt_0)}\right\rceil-1\geq0$, with $2(2+\wt_0)<\gt_0<2+\wt_0$, is the maximum number of terms in the series. Notice that for each $m$-order of the expansion is valid if $m(\gt_0-2(2+\wt_0))<2+\wt_0-\gt_0$.

\subsubsection{case of $\st_0=0$}

In this case the expansions (\ref{couplings expansion}) becomes
\begin{eqnarray}
  w&=&w_0 f(r)^{\wt_0}+\cdots,\qquad G=g_0 f(r)^{\gt_0}+\cdots,\qquad  S=2b^2+ s_1 f(r)^{\st_1}+\cdots,
\end{eqnarray}
where we have taken $s_0=2b^2$, with $s_1\neq0$ and $\st_1>0$ are constant and finite. The possible leading terms of the constraint equation (\ref{CE4}) are
\begin{equation}
 \propto 4 b^4 g_0^2 s_1^2 w_0 (\gt_0+\st_1)^2 f(r)^{\gt_0+\st_1}+32 b^4
   g_0 s_1 w_0^3 f(r)^{2 \wt_0+4}+\cdots.
\end{equation}
The leading terms will be zero if $\gt_0=2(2+\wt_0)-\st_1$ and $g_0=-{2 w_0^2 \over (2+\wt_0)^2s_1}$. Following the same steps as in the previous case we obtain the expansion of $G$ as follows:
\begin{equation}
 G=-{2 w_0^2 \over s_1(2+\wt_0)^2}f(r)^{4+2\wt_0-\st_1}\sum^{m_{max}}_{m=0}\left(-{s_1 \over 2b^2} f(r)^{\st_1}\right)^m+\cdots,
\end{equation}
where for each $m$-th order of the expansion $2+\wt_0+m~\st_1<0$. Similar to the previous case, since $\st_1>0$, there is higest order of expansion at $m_{max}=-\left\lceil{2+\wt_0\over \st_1}\right\rceil-1$, which means the expansion of $G$ is exact. We could also make expansion of $S$ by taking $\st_1=2(2+\wt_0)-\gt_0$ and $s_1=-{2 w_0^2 \over (2+\wt_0)^2g_0}$ instead. As a resuls the expansion of $S$ is
\begin{equation}
 S=2b^2\left(1- {w_0^2f^{4+2\wt_0-\gt_0}\over b^2 g_0(2+\wt_0)^2}-{w_0^4 f^{2(4+2\wt_0-\gt_0)} \over b^4 g_0^2 (2+\wt_0)^4}-{2w_0^6 f^{3(4+2\wt_0-\gt_0)}\over b^6 g_0^3 (2+\wt_0)^6}-{5w_0^8 f^{4(4+2\wt_0-\gt_0)}\over b^8 g_0^4 (2+\wt_0)^8}+\cdots\right).
\end{equation}
The second term on the right hand side is valid if $4+2\wt_0-\gt_0>0$, or $\gt_0<0$ since $2+\wt_0<0$, while the third term is valid if $3(2+\wt_0)-\gt_0<0$. The fourth and five terms are valid if $5(2+\wt_0)-2\gt_0<0$ and $7(2+\wt_0)-3\gt_0<0$ respectively. Similarly like in the case of $\st_0<0$, the expansion of $S$ could be simply written as
\begin{equation}
 S=2b^2-{2w_0^2f^{4+2\wt_0-\gt_0}\over g_0(2+\wt_0)^2} \sum^{m_{max}}_{m=0}\textbf{C}_m\left(w_0^2 f^{4+2\wt_0-\gt_0}\over b^2 g_0(2+\wt_0)^2\right)^m,
\end{equation}
where for each $m$-order of the expansion is valid if $2+\wt_0<m (\gt_0-2(2+\wt_0))$ except for $m=0$ which is valid if $4+2\wt_0-\gt_0>0$ or means $\gt_0<0$. Since $\gt_0<0$, then there is maximum value of $m$ at $m_{max}=\left\lceil{2+\wt_0\over \gt_0-2(2+\wt_0)}\right\rceil-1$. Hence the expansion of $S$ is exact as well.

If $s_0\neq 2b^2$ then we have $g_0=-{4b^2 w_0^2 \over s_0(s_0-2b^2)(2+\wt_0)^2}$ and $\gt_0=4+2\wt_0$. So the expansions (\ref{couplings expansion}) can be rewritten as follows
\begin{eqnarray}
  w&=&w_0 f(r)^{\wt_0}+\cdots,~ G=-{4b^2 w_0^2 \over s_0(s_0-2b^2)(2+\wt_0)^2} f(r)^{4+2\wt_0}+ g_1 f(r)^{4+2\wt_0+\gt_1}\cdots,\nonumber\\
  S&=&s_0 +s_1 f(r)^{\st_1}+\cdots,
\end{eqnarray}
with $g_1\neq0$; $s_1\neq0$; $\gt_1>0$; and $\st_1>0$ are constant and finite. The leading term then is given by
\begin{equation}
 \propto \frac{128~b^4  s_1 w_0^5 \left(b^2-s_0\right)\st_1}{s_0 (\wt_0+2)^3 \left(2 b^2-s_0\right)}.
\end{equation}
This will be the vanishing leading term if $s_1=0$ or $s_0=b^2$. Unfortunately both will lead to $C_0=0$ or $w_0=0$ which is not allowed.

\subsubsection{case of $\st_0>0$}
The expansions (\ref{couplings expansion}) can be writen as
\begin{eqnarray}
  w&=&w_0 f(r)^{\wt_0}+\cdots,\qquad G={2w_0^2 \over (2+\wt_0)^2 s_0}f(r)^{4+2\wt_0-\st_0}+g_1 f(r)^{4+2\wt_0-\st_0+\gt_1}+\cdots,\nonumber\\
  S&=&s_0 f(r)^{\st_0}+\cdots,
\end{eqnarray}
with $g_1\neq0$ and $\gt_1>0$ are constant and finite. It turns out that the expansion of $G$ can be expressed as
\begin{equation}
 G={2w_0^2 \over (2+\wt_0)^2 s_0}f(r)^{4+2\wt_0-\st_0} \sum^{m_{max}}_{m=0}\left(s_0 f^{\st_0}\over 2 b^2 \right)^m,
\end{equation}
where for each $m$-order of the expansion is valid if $2+\wt_0+m~\st_0<0$. Therefore the maximum $m$-order of expansion is at $m_{max}=-\left\lceil{2+\wt_0\over \st_0}\right\rceil-1$, which means the expansion of $G$ is exact. Similarly we could also find expansion of $S$ instead by using the following expansions
\begin{eqnarray}
  w&=&w_0 f(r)^{\wt_0}+\cdots,\qquad G=g_0f(r)^{\gt_0}+\cdots,\nonumber\\
  S&=&{2w_0^2 \over (2+\wt_0)^2 g_0} f(r)^{4+2\wt_0-\gt_0}+s_1 f(r)^{4+2\wt_0-\gt_0+\st_1}+\cdots,
\end{eqnarray}
where $\gt_0<4+2\wt_0<0$; $s_1\neq0$; and $\st_1>0$ are constant and finite. The expansion is given by
\begin{eqnarray}
 S&=&{2w_0^2 f(r)^{4+2\wt_0-\gt_0}\over (2+\wt_0)^2 g_0} \left(1+{w_0^2 f(r)^{4+2\wt_0-\gt_0}\over b^2 g_0 (2+\wt_0)^2}+{2 w_0^4 f(r)^{2(4+2\wt_0-\gt_0)} \over b^4 g_0^2 (2+\wt_0)^4}+{5 w_0^6 f(r)^{3(4+2\wt_0-\gt_0)}\over b^6 g_0^3 (2+\wt_0)^6}\right.\nonumber\\
 &&\left.+{14 w_0^8 f(r)^{4(4+2\wt_0-\gt_0)} \over b^8 g_0^4 (2+\wt_0)^8}+\cdots\right).
\end{eqnarray}
As before, we would like to write the expansion of $S$ in a more simple form as
\begin{equation}
 S={2w_0^2 f(r)^{4+2\wt_0-\gt_0}\over (2+\wt_0)^2 g_0} \sum^{m_{max}}_{m=0} \textbf{C}_m \left(w_0^2 f(r)^{4+2\wt_0-\gt_0} \over b^2 g_0 (2+\wt_0)^2\right)^m,
\end{equation}
where for each $m$-order of the expansion is valid if $2+\wt_0<m (\gt_0-2(2+\wt_0))$ except for $m=0$ which is valid if $4+2\wt_0-\gt_0>0$, or $\gt_0<0$. Since $\gt_0<0$, then there is maximum value of $m$ at $m_{max}=\left\lceil{2+\wt_0\over \gt_0-2(2+\wt_0)}\right\rceil-1$ which means the expansion of $S$ is also exact.

As summary, we can rewrite the expansions of $G$ and $S$, with $\wt_0<-2$, simply in terms of few constants as follows
\begin{description}
 \item[\underline{$\st_0<0$}:]
 \begin{equation}
 G=-{4b^2 w_0^2 \over s_0^2 (2+\wt_0)^2}f(r)^{2(2+\wt_0-\st_0)}\sum^{g_{max}}_{m_g=0}\left({2 b^2 \over s_0} f(r)^{-\st_0}\right)^{m_g},
\end{equation}
\begin{eqnarray}
 S&=& s_0 f(r)^{\st_0}+b^2+ {b^4 \over 2 s_0}f(r)^{-\st_0}\sum^{s_{max}}_{m_s=0}\textbf{C}_{ms}\left(-{b^4 f(r)^{-2\st_0} \over 4 s_0^2}\right)^{m_s},
\end{eqnarray}
with $g_{max}=\left\lceil{2+\wt_0\over \st_0}\right\rceil-1\geq0$ and $s_{max}=\left\lceil{2+\wt_0\over 2\st_0}\right\rceil-2\geq0$. The value of $\st_0$ must satisfy $2+\wt_0<g_{max}~\st_0$ and $(2+\wt_0)<2\st_0(s_{max}+1)$. If $s_{max}<0$ then the expansion of $S$ is valid up to second term on the right hand side if only if $2+\wt_0<\st_0$.

 \item[\underline{$\st_0=0$}:]
 \begin{equation}
 G=-{2 w_0^2 \over s_1(2+\wt_0)^2}f(r)^{4+2\wt_0-\st_1}\sum^{m_{max}}_{m=0}\left(-{s_1 \over 2b^2} f(r)^{\st_1}\right)^m,
\end{equation}
\begin{equation}
 S=2b^2+s_1 f(r)^{\st_1} \sum^{m_{max}}_{m=0}\textbf{C}_m\left(-{s_1 f(r)^{\st_1}\over 2b^2}\right)^m,
\end{equation}
with $m_{max}=-\left\lceil{2+\wt_0\over \st_1}\right\rceil-1\geq0$. The value of $\st_1$ must satisfy $m_{max}\st_1<-(2+\wt_0)$ except for $m_{max}=0$ then it must satisfy $\st_1>0$.

 \item[\underline{$\st_0>0$}:]
 \begin{equation}
 G={2w_0^2 \over (2+\wt_0)^2 s_0}f(r)^{4+2\wt_0-\st_0} \sum^{m_{max}}_{m=0}\left(s_0 f(r)^{\st_0}\over 2 b^2 \right)^m,
\end{equation}
\begin{equation}
 S=s_0 f(r)^{\st_0} \sum^{m_{max}}_{m=0} \textbf{C}_m \left(s_0 f(r)^{\st_0} \over 2b^2\right)^m,
\end{equation}
with  $m_{max}=-\left\lceil{2+\wt_0\over \st_0}\right\rceil-1\geq0$. The value of $\st_0$ must satisfy $m_{max}~\st_0<-(2+\wt_0)$.
\end{description}

For futher analysis we will choose those expansions in which $2+\wt_0<\st_0$, for $\st_0<0$; $-\st_1>2+\wt_0$, for $\st_0=0$; and $-\st_0>2+\wt_0$, for $\st_0>0$. Therefore the near origin expansions are
\begin{itemize}
 \item[\underline{$\st_0<0$}:]
 \begin{equation}
 G=-{4b^2 w_0^2 \over s_0^2 (2+\wt_0)^2}f(r)^{2(2+\wt_0-\st_0)}\left(1+{2 b^2 \over s_0} f(r)^{-\st_0}\right).
\end{equation}
\begin{eqnarray}
 S&=& s_0 f(r)^{\st_0}+b^2.
\end{eqnarray}

\item[\underline{$\st_0=0$}:]
 \begin{equation}
 G=-{2 w_0^2 \over s_1(2+\wt_0)^2}f(r)^{4+2\wt_0-\st_1}\left(1-{s_1 \over 2b^2} f(r)^{\st_1}\right),
\end{equation}
\begin{equation}
 S=2b^2+s_1 f(r)^{\st_1}\left(1-{s_1 f(r)^{\st_1}\over 2b^2}\right).
\end{equation}

 \item[\underline{$\st_0>0$}:]
 \begin{equation}
 G={2w_0^2 \over (2+\wt_0)^2 s_0}f(r)^{4+2\wt_0-\st_0} \left(1+s_0 f(r)^{\st_0}\over 2 b^2 \right),
\end{equation}
\begin{equation}
 S=s_0 f(r)^{\st_0} \left(1+s_0 f(r)^{\st_0} \over 2b^2\right).
\end{equation}
\end{itemize}
Substituting these expansions into the BPS equations (\ref{BPS BIH a}) and (\ref{BPS BIH f}) we get near origin expansions of $a$  and $f$ as follows\footnote{There are actually more possible expansions of $f$ in the case of $\st_0<0$ that depend on the value of $n\st_0+2$. Here we take only the one with $n\st_0<-2$ that most likely valid for arbitrary values of $n>0$.}:
\begin{itemize}
 \item [\underline{$\st_0<0$}:]
 \begin{eqnarray}
  a(r)&=&n\pm{f0^{-2 + \st_0 - \wt_0} s_0 (2 + \wt_0)\over(-4 + 2 n (2 - \st_0 + \wt_0)) w_0}~ r^{2 + n (-2 + \st_0 - \wt_0)}+\cdots,\\
  f(r)&=& f_0 r^n\mp{f_0^{\st_0 - 1-\wt_0} s_0 (2 + \wt_0)\over
 2 (2 - n (2 + \wt_0 - \st_0))^2 w_0} r^{2 + n(\st_0 - 1-\wt_0)}+\cdots,
 \end{eqnarray}

 \item [\underline{$\st_0=0$}:]
 \begin{eqnarray}
  a(r)&=&n\pm{f0^{-2 + \st_1 - \wt_0} s_1 (2 + \wt_0)\over (2 - n (2 - \st_1 + \wt_0)) w_0}~ r^{2 + n (-2 + \st_1 - \wt_0)}+\cdots,\\
  f(r)&=& f_0 r^n-\frac{C_0 f_0^{-\wt_0-1}}{2 w_0(2+\wt_0)} r^{-n(1+\wt_0)}+\cdots,
 \end{eqnarray}
 
 \item [\underline{$\st_0>0$}:]
 \begin{eqnarray}
 a(r)&=& n\pm\frac{f_0^{-2+\st_0-\wt_0}s_0 (2+\wt_0) }{2-(n (2-\st_0+\wt0-2))w_0 } r^{2 + n (-2+\st_0 -\wt_0)}+\cdots,\\
 f(r)&=& f_0 r^n-\frac{C_0 f_0^{-\wt_0-1}}{2 w_0(2+\wt_0)} r^{-n(1+\wt_0)}+\cdots,
\end{eqnarray}
with $n>0$.
\end{itemize}

\subsection{Near boundary expansions}
Near the boundary, $f(r\to\infty)\to 1$, the expansion of $w$ can be written as
\begin{equation}
 w=w_{b0} (1-f(r))^{\wt_{b0}}+\cdots,
\end{equation}
with $w_{b0}\neq0$ and $\wt_{b0}$ are constant and finite, while the expansions of $a$ and $f$ are
\begin{eqnarray}
 a(r)&=&a_{b0}~ r^{-\at_{b0}}+\cdots,\\
 f(r)&=&f_{b0}~ r^{-\ft_{b0}}+\cdots,
\end{eqnarray}
with $a_{b0}\neq0$; $f_{b0}\neq0$; $\at_{b0}>0$ and $\ft_{b0}>0$ are constant and finite.

\subsubsection{case of $\st_0<0$}
In this case we take the expansions of $G$ and $S$ as follows
\begin{eqnarray}
 G&=&-\frac{4 b^2
   w_0^2 \left(2 b^2+s_0\right)}{s_0^3 (\wt_0+2)^2}+\frac{8 b^2 w_0^2 \left(b^2 (-3 \st_0+2
   \wt_0+4)+s_0 (-\st_0+\wt_0+2)\right)}{s_0^3 (\wt_0+2)^2}(1-f(r))\nonumber\\
   &&-\frac{4 b^2 w_0^2 \left(b^2 (3 \st_0-2 \wt_0-3) (3 \st_0-2
   (\wt_0+2))+s_0 (2 \st_0-2 \wt_0-3) (\st_0-\wt_0-2)\right)}{s_0^3
   (\wt_0+2)^2}(1-f(r))^2\nonumber\\
   &&+\cdots \\
   S&=&b^2+s_0-\st_0 s_0 (1-f(r))+\frac{1}{2} (\st_0-1) \st_0
   s_0 (1-f(r))^2-\frac{1}{6} (\st_0-2) (\st_0-1) \st_0 s_0 (1-f(r))^3\nonumber\\
   &&+\cdots.
\end{eqnarray}
Substituting these into the constraint equation (\ref{CE4}) gives us three possible values of $s_0$ which are $-2b^2,-b^2$ and $b^2$. For $s_0=-2b^2$, the expansion of $w$ becomes
\begin{equation}
 w=\pm {w_0\over 4(2+\wt_0)}\sqrt{-3\st_0}(1-f(r))^{-1/2}+{C_0\over 2}+\cdots.
\end{equation}
However, the possible leading terms of the BPS equation (\ref{BPS BIH a}),
\begin{equation}
 \propto 4 \at_{b0}^2 a_{b0}^2 w_0^2+3 b^4 (\wt_0+2)^2 r^{2
   \at_{b0}+4},
\end{equation}
tells us that $\at_{b0}=-2<0$ which is not allowed. For $s_0=\pm b^2$, there are two possible expansions of $w$ which are
\begin{eqnarray}
 w&=&\pm {w_0\over (2+\wt_0)}\sqrt{-3\st_0\over 2}(1-f(r))^{-1/2}+{C_0\over 2}+\cdots,
\end{eqnarray}
which is only valid for $s_0=b^2$, and
\begin{eqnarray}
 w&=&-C_0+w_{b1} (1-f(r))+\cdots,
\end{eqnarray}
with $w_{b1}=-2 C_0 - {2 C_0^3 (2 + \wt_0)^2)\over(3 \st_0 w_0^2}$, for $s_0=b^2$, and $w_{b1}=-2 C_0 +{2 C_0^3 (2 + \wt_0)^2)\over(\st_0 w_0^2}$, for $s_0=-b^2$.
Substituting the expansion of $G$ and $S$ above into the BPS equation (\ref{BPS BIH a}), we obtain $\ft_{b0} = 4 + 2~ \at_{b0}$. However, substituting those into the BPS equation (\ref{BPS BIH f}) gives us $\ft_{b0} = \at_{b0}$ and combining with the previous result implies $\at_{b0}=-4<0$ which is not allowed.

\subsubsection{case of $\st_0=0$}
For this case we will use the following expansions of $G$ and $S$:
\begin{eqnarray}
 G&=&\frac{w_0^2 \left(s_1-2 b^2\right)}{b^2
   s_1 (\wt_0+2)^2}-\frac{2 w_0^2 
   \left(b^2 (\st_1-2 (\wt_0+2))+s_1 (\wt_0+2)\right)}{b^2 s_1
   (\wt_0+2)^2}(1-f(r))\nonumber\\
   &&+\frac{w_0^2 \left(s_1 (\wt_0+2) (2 \wt_0+3)-b^2 (\st_1-2 \wt_0-3)
   (\st_1-2 (\wt_0+2))\right)}{b^2 s_1 (\wt_0+2)^2}(1-f(r))^2\nonumber\\
   &&+\cdots \\
   S&=&-\frac{s_1^2}{2
   b^2}+2 b^2+s_1+\st_1 s_1
   \left(\frac{s_1}{b^2}-1\right) (1-f(r))+\frac{\st_1 s_1 \left(b^2
   (\st_1-1)-2 \st_1 s_1+s_1\right)}{2 b^2}(1-f(r))^2\nonumber\\
   &&-\frac{(\st_1-1) \st_1 s_1 \left(b^2 (\st_1-2)+(2-4 \st_1)
   s_1\right)}{6 b^2}(1-f(r))^3+\cdots.
\end{eqnarray}
Again substituting these expansions into the constraint equation (\ref{CE4}) gives us three possible values of $s_1$ which are $2b^2, (1+\sqrt{5})b^2,$ and $(1-\sqrt{5})b^2$. For $s_1=2b^2$, the expansion of $w$ becomes
\begin{equation}
 w=-{3\over 2}C_0-{3\over 2}(7+2\wt_0)C_0 (1-f(r))+\cdots,
\end{equation}
where now $C_0=\pm{2 \st_1 w_0\over 3 \sqrt{3} (2 + \wt_0)}$ is not arbitrary. If we substitute the above expansion of $G$ and $S$, with $s_1=2b^2$, we obtain the possible leading terms which are given by
\begin{equation}
 \propto 4 b^4  (2 + \wt_0)^2 r^{4 + 2 \at_{b0}} - \at_{b0}^2 a_{b0}^2 w_0^2.
\end{equation}
This implies $\at_{b0}=-2<0$ which is not allowed. For $s_1=(1\pm\sqrt{5})b^2$, there are two possible expansions of $w$ which are
\begin{equation}
 w=\pm{w_0\over 2 + \wt_0} \sqrt{{\st_1\over 8 b^2}\left(6 b^2 - s_1\right)}(1-f(r))^{-1/2}+{C_0\over 2}+\cdots
\end{equation}
and
\begin{equation}
 w=-C_0+\left(-2 C_0 + {8 C_0^3 (2 + \wt_0)^2 b^2 \over \left(s_1 - 6 b^2\right) \st_1 w_0^2}\right)(1-f(r))+\cdots.
\end{equation}
Substituting the expansions of $G$ and $S$, with $s_1=(1\pm\sqrt{5})b^2$, into the BPS equation (\ref{BPS BIH a}) gives us $\ft_{b0}=4+2\at_{b0}$, while into the BPS equation (\ref{BPS BIH f}) gives us $\ft_{b0}=\at_{b0}$ or $\ft_{b0}=2\at_{b0}$ respectively for different expansions of $w$ above. Both expansions of $w$ leads to $\at_{b0}=-4<0$ or contradiction, respectively, and thus are not allowed.

\subsubsection{case of $\st_0>0$}
In this last case, the expansions of $G$ and $S$ can be written as
\begin{eqnarray}
 G&=&\frac{w_0^2 \left(2 b^2+s_0\right)}{b^2
   s_0 (\wt_0+2)^2}-\frac{2 w_0^2
    \left(b^2 (-\st_0+2 \wt_0+4)+s_0 (\wt_0+2)\right)}{b^2 s_0
   (\wt_0+2)^2}(1-f(r))\nonumber\\
   &&+\frac{w_0^2  \left(b^2 (\st_0-2 \wt_0-3) (\st_0-2 (\wt_0+2))+s_0
   (\wt_0+2) (2 \wt_0+3)\right)}{b^2 s_0 (\wt_0+2)^2}(1-f(r))^2+\cdots,\nonumber\\
   \\
   S&=&\frac{s_0^2}{2
   b^2}+s_0-\frac{\st_0 s_0
   \left(b^2+s_0\right)}{b^2} (1-f(r))+\frac{\st_0 s_0 \left(b^2
   (\st_0-1)+(2 \st_0-1) s_0\right)}{2 b^2}(1-f(r))^2 \nonumber\\
   &&-\frac{(\st_0-1) \st_0 s_0 \left(b^2 (\st_0-2)+4 \st_0 s_0-2
   s_0\right)}{6 b^2}(1-f(r))^3+\cdots.
\end{eqnarray}
Similarly substituting these expansions into the constraint equation (\ref{CE4}) give us three possible values of $s_0$ which are $-2b^2, (-1-\sqrt{5})b^2,$ and $(-1+\sqrt{5})b^2$. For $s_0=-2b^2$, the expansion of $w$ is
\begin{equation}
 w=-{3\over 2}C_0-{3\over 2}(7+2\wt_0)C_0 (1-f(r))+\cdots,
\end{equation}
where now $C_0=\pm{2 \st_1 w_0\over 3 \sqrt{3} (2 + \wt_0)}$ is not arbitrary. Unfortunately if we substitute the expasions of $G$ and $S$, with $s_0=-2b^2$, into the BPS equation (\ref{BPS BIH a}) the leading terms,
\begin{equation}
 \propto 4 \at_{b0}^2 a_{b0}^2 w_0^2+3 b^4 (\wt_0+2)^2 r^{2
   \at_{b0}+4},
\end{equation}
imply $a=-2<0$ which is not allowed. For $s_0=(-1\pm\sqrt{5})b^2$, the possible expansions of $w$ are
\begin{equation}
 w=\pm {w_0 \over 2 (2 + \wt_0)} \sqrt{\st_0 (s_0 + 6 b^2)\over 2 b^2} (1-f(r))^{-1/2}+{C_0\over 2}+\cdots
\end{equation}
and 
\begin{equation}
 w=-C_0+\left(-2 C_0 + {2 C_0^3 (2 + \wt_0)^2 (s_0 - 4 b^2)\over 5 \st_0 w_0^2 b^2}\right)(1-f(r))+\cdots.
\end{equation}
Substituting the expansions of $G$ and $S$ into the BPS equation (\ref{BPS BIH a}) gives us $\ft_{b0}=4+2\at_{b0}$, while into the BPS equation (\ref{BPS BIH f}) gives us $\ft_{b0}=\at_{b0}$ and $\ft_{b0}=2\at_{b0}$ respectively for different expansions of $w$ above. Both expansions of $w$ leads to $\at_{b0}=-4<0$ or contradiction, respectively, and thus are not allowed.

% We also need to check if first derivative of potensial over scalar field near boundary is zero, $\lim_{r\to\infty} {\partial V \over \partial f}\to 0$. Using the constraint equations (\ref{CE4}) and (\ref{constraint X0}), we obtain
% \begin{equation}
%  {\partial V \over \partial f}=-{S(2b^2-S) \over 2~G(b^2-S)}{\partial G\over \partial f}+{2~f\over (b^2-S)}\sqrt{b^2w~S(2b^2-S)(f^2w+C_0)\over G}.
% \end{equation}
% Since the near boundary also corresponds to $f(r)\to1$, we will use the following expansions of $G$ and $S$:

\section{Conclusions and Remarks}
\label{sec 6}
In this article we have shown how to use BPS Lagrangian method to derive BPS equations for vortices with nonzero stress tensor in the generalized Maxwell-Higgs(\ref{Gen Lagrangian}) and the generalized Born-Infeld-Higgs models (\ref{Lagrangian GenBIH}). In order to get BPS equations for vortices with nonzero stress tensor we added terms into the initial BPS Lagrangian, which is a total derivative term $\mathcal{L}=\int dQ$, by two additional terms proportional to first-derivative of the scalar effective field, $(f')^0$ and $(f')^2$, multiplied respectively by arbitrary functions $X_0$ and $X_2$ that in general depend only on the effective fields and radial coodinate. However, being not total derivative terms, these additional terms give additional constraint equations resulting from Euler-Lagrange equations of the BPS Lagrangian. This implies larger number of constraint equations than the number of arbitrary functions in the BPS Lagrangian that needs to be solved. We took several assumptions on the functions $X_0$ and $X_2$ in order to match the number of constraint equations and the number of arbitrary functions in the BPS Lagrangian as close as possible. At the end, when we could not take further assumption, the resulting number of constraint equations is one larger than the number of arbitrary functions in the BPS Lagrangian. We found that the additional terms in both models, (\ref{Gen Lagrangian}) and (\ref{Lagrangian GenBIH}), are similarly given by
\begin{equation}
 \CL_{BPS}=-{Q'(f)\over r} f'-{Q'(a)\over r}a'-X_0(f)-X_2(f) f'^2,
\end{equation}
where $Q\equiv Q(a,f)$.  Both the arbitrary functions $X_0$ and $X_2$ are related to each other by means $X_0=0$ if and only if $X_2=0$. Furthermore, in both models, the solutions for $X_2$ are equally given by (\ref{Sol X_2}) and the constraint equation for $X_0$ has the same form of (\ref{constraint X0}). The finite energy condition implied a constraint on integral of $X_0$ over the whole space given by (\ref{int X_0}). There is still one remaining constraint equation in both models, which are (\ref{CE3}) and (\ref{CE4}) respectively, that would fix one of the arbitrary functions ($w,G,V$) in their effective Lagrangians\footnote{In their non-generalized models this constraint equation will fix the form of scalar potential $V$.}. We also tried to consider more general BPS Lagrangian by adding a term that is proportional to $(a')^2$, multiplied by an arbitrary function $X_1(a,f;r)$, and it turns out this term should be vanishing. The total static energy of the BPS equations of both models is finite. The finite energy condition constraints the values of $V$,
\begin{eqnarray}
 2b^2> V >0
\end{eqnarray}
and $V\neq b^2$. Requirement that the BPS equations (\ref{BPS BIH a}) and (\ref{BPS BIH f}) must be real and well defined give additional bounds to the values of $X_0$,
\begin{equation}
 2b^2>V-X_0>0
\end{equation}
and $V-X_0\neq b^2$. Those bounds are also valid for the generalized Maxwell-Higss model with $b^2\to\infty$.
We later computed the energy momentum tensor and showed explicitly diagonal components of the stress tensor in radial and angular directions, $T_{rr}$ and $T_{\theta\theta}$ repectively, took the same forms of (\ref{Tradial}) and (\ref{Tangle}) in the BPS limit for both models. Both $T_{rr}$ and $T_{\theta\theta}$ can not be zero independently otherwise it will lead to a constant value of $a$ by the constraint equation (\ref{constraint X0}), but they can be simultaneously zero if $X_0=X_2=0$ or $C_0=0$. Therefore nonzero in all diagonal components of the stress tensor means the resulting BPS vortices exhibit forces in all directions. The last but not least, we computed the conservation of energy-momentum tensor and turns out it is the same equation as the constraint equation (\ref{constraint X0}) and hence no additional constraint equations need to be considered.

We did the numerical analysis for possible functions of $w,G,$ and $S\equiv V-X_0$ that could satisfy the constraint equation (\ref{CE4}). From the near origin analysis, $f(r\to0)\to0$, of the BPS equation (\ref{BPS BIH f}) we found the leading order of $w$ must be proportional to $f(r)^{\wt_0}$, with $\wt_0<-2$, and it turned out that the near origin expansions of $G$ and $S$ are exact. We then considered a large class of solutions of $G$ and $S$ and unfortunately we found that the effective fields $a$ and $f$ diverge near the boundary at the order of $O(r^4)$. Therefore we may conclude that the BPS equations (\ref{BPS BIH a}) and (\ref{BPS BIH f}), with the constraint equations (\ref{CE4}) and (\ref{constraint X0}), do not have solutions that behave nicely near the boundary. We suggest that incorporating gravity, namely  adding the Einstein-Hilbert action, into the Born-Infeld-Higgs action may resolve this near boundary problem. As we have shown previously that this type of BPS vortices exhibit some forces due to nonzero value of the stress tensor and thus they may not exist as stable solutions. We expect that gravitational force would overcome these forces. However, this also means we have to improve the BPS Lagrangian method for obtaining BPS equations for the gravitational field and this would be investigated further in the next article. Other possible resouliton could also come from considering higher dimensional models and find the corresponding BPS equations for vortices. As an example one could consider models in four dimensional spacetime and find BPS equations for string vortices. In the spherical coordinates, the effective action has additional factor of $r^1$ from its Jacobian compared the one in three dimensional models and it that would modify the BPS equations and the constraint equations. We expect that they could have solutions that behave nicely near the boundary. Once the BPS vortices found it might be interesting to study the supersymmetric realization of those vortices.

From the two models studied in this article, we may learn that BPS equations for vortices with zero stress tensor are due to BPS Lagrangian having only the total derivative terms,
\begin{equation}
 L_{BPS}=\int d^2x~\CL_{BPS}=\int dQ\propto\int d^2x\left(-{Q'(f)\over r} f'-{Q'(a)\over r}a'\right).
\end{equation}
The BPS equations for vortices with nonzero stress tensor can be constructed from the coresponding BPS Lagrangian by adding more terms, which are not total derivative terms, into it as shown previously. In this fashion, we expect that one could find BPS equations for other type of solitons with nonzero stress tensor in other models and dimensions. However this may not be the only way to get BPS equations for vortices with nonzero stress tensor. As an example one could consider BPS Lagrangian for the case of generalized Maxwell-Higgs model as follows
\begin{equation}
 \CL_{BPS}=X_0(a,f,r)+X_{a1}(a,f,r) a'+X_{f1}(a,f,r) f',\label{other BPS Lagragian}
\end{equation}
where now all the arbitrary functions are functions of all the effective fields and of (radial) coordinate explicitly.
In this case, the BPS equations are given by
\begin{eqnarray}
a'&=&-{r^2 \over G}X_{a1},\\
f'&=&-{1\over 2 w}X_{f1},
\end{eqnarray}
which imply $X_{a1}\neq0$ and $X_{f1}\neq0$ for notrivial solutions. In addition there are three constraint equations:
\begin{equation}
\left({r^2 X_{a1}^2\over 2G}-V-X_0\right)+w\left({X_{f1}^2 \over 4 w^2}-{a^2f^2\over r^2}\right)=0,\label{cons 1}
\end{equation}
the two others are derived from the Euler-Lagrange of its BPS Lagrangian,
\begin{equation}
 {X_{a1}\over r}={\partial X_0 \over \partial a}-{\partial X_{a1} \over \partial r}-\left({\partial X_{f1} \over \partial a}-{\partial X_{a1} \over \partial f}\right){X_{f1}\over 2w}\label{cons 2}
\end{equation}
and
\begin{equation}
 {X_{f1}\over r}={\partial X_0 \over \partial f}-{\partial X_{f1} \over \partial r}-\left({\partial X_{a1} \over \partial f}-{\partial X_{f1} \over \partial a}\right){r^2 X_{a1}\over G}.\label{cons 3}
\end{equation}
Diagonal components of the stress tensor in radial and angular directions are given by
\begin{eqnarray}
 T_{rr}&=&{r^2 X_{a1}^2\over 2G}+w\left({X_{f1}^2 \over 4 w^2}-{a^2f^2\over r^2}\right)-V,\\
 T_{\theta\theta}&=&r\left({r^2 X_{a1}^2\over 2G}-w\left({X_{f1}^2 \over 4 w^2}-{a^2f^2\over r^2}\right)-V\right),
\end{eqnarray}
where we have taken the BPS limit. Notice that the constraint equation (\ref{cons 1}) can be recasted into $T_{rr}=X_0$. First, let us consider BPS equation for vortices with zero stress tensor by imposing ``pressureless'' condition, $T_{rr}=T_{\theta\theta}=0$, as such
\begin{equation}
 X_{f1}=\pm2w{af\over r},\qquad\qquad X_{a1}=\pm{\sqrt{2GV}\over r},\label{Sol X0=0}
\end{equation}
which then implies $X_0=0$ by using the constraint equation (\ref{cons 1}). Furthermore, using either the constraint equation (\ref{cons 2}) or (\ref{cons 3}), we obtain ${\partial X_{a1} \over \partial f}={\partial X_{f1} \over \partial a}$, or ${\partial\sqrt{2GV}\over\partial f}=2wf$, which turns the last two terms in the BPS Lagrangian (\ref{other BPS Lagragian}) to become total derivative terms. For general cases, where $X_0$ does not have to be zero, the number of arbitrary functions in the BPS Lagrangian is equal to the number of constraint equations that is three. So there could be other BPS equations for vortex with nonzero stress tensor, using  the BPS Lagrangian (\ref{other BPS Lagragian}), besides the ones that we have discussed previously. Further simplification by substituting $X_0$ from equation (\ref{cons 1}) into equations (\ref{cons 2}) and (\ref{cons 3}) yields two constraint equations:
\begin{equation}
 {dX_{a1}\over dr}+{X_{a1}\over r}=- {2\over r^2}awf^2
\end{equation}
and
\begin{equation}
  {dX_{f1}\over dr}+{X_{f1}\over r}=- \left({r^2 X^2_{a1}\over 2G^2}{\partial G\over \partial f}+{X^2_{f1}\over 4w^2}{\partial w\over\partial f}\right)-{a^2\over r^2}{\partial(f^2 w)\over\partial f}-{\partial V\over\partial f}.
\end{equation}
One can verifiy that these constraint equations are also satisfied by solutions of $X_0=0$, which are given by (\ref{Sol X0=0}) along with a constraint equation ${\partial\sqrt{2GV}\over\partial f}=2wf$. Solutions of these constraint equations other than (\ref{Sol X0=0}) implies nonzero stress tensor in radial and/or angular directions.

\acknowledgments
We would like to thank the ICTP Associate Scheme office for the support and hospitality during our visit to the Abdus Salam ICTP, under the ICTP Associateship Scheme programme, where the finishing of this article was done. We also would like to thank Edi Gava for useful comments and suggestions.

\bibliography{testBiB}
\bibliographystyle{hieeetr}
\end{document}